\documentclass{PoS}

\rightline{\sffamily UTCCS-P-32, UTHEP-550}\vspace*{-10mm}

\title{Marking up lattice QCD configurations and ensembles}

\ShortTitle{Marking up lattice QCD configurations and ensembles}

\author{P.~Coddington\\
        School of Computer Science, University of Adelaide,
        Adelaide, 5005, Australia}
\author{B.~Jo\'o\\
        Jefferson Lab, Newport News, VA 23606, USA}
\author{C.~M.~Maynard\\
        EPCC, School of Physics, University of Edinburgh,
        Edinburgh EH9 3JZ, UK}
\author{D.~Pleiter\\
        Deutsches Elektronen-Synchrotron DESY, 15738 Zeuthen, Germany}
\author{\speaker{T.~Yoshi\'e}%
        \thanks{E-mail: yoshie@ccs.tsukuba.ac.jp}\\
         Center for Computational Sciences, University of Tsukuba,
         Tsukuba 305-8577, Japan}
\author{ILDG Metadata Working Group}

\abstract{
QCDml is an XML-based markup language designed for sharing QCD
configurations and ensembles world-wide via the International Lattice
Data Grid (ILDG). 
Based on the latest release, we present key ingredients of the QCDml
in order to provide some starting points for colleagues in this
community to markup valuable configurations and submit them to the ILDG.
}

\FullConference{The XXV International Symposium on Lattice Field Theory\\
		 July 30 - August 4 2007\\
		 Regensburg, Germany}

\begin{document}

\section{Introduction}
Sharing QCD configurations world-wide via 
the ILDG~\cite{ref:ILDG2002,ref:ILDG2003,ref:ILDG20046,ref:ILDG2007}
requires standards of notation and terminology for metadata of
configurations such as lattice actions and sizes used in simulations. 
For this purpose, the Metadata Working Group 
(MDWG)~\cite{ref:MDWGmember} has been 
working~\cite{ref:ILDG2003,ref:MDWG2004,ref:MDWG2006} on
designing an XML-based markup language QCDml~\cite{ref:ILDGweb}, 
a set of rules for XML instance documents (IDs). 
IDs are stored in regional grid databases which are
aggregated together to form the ILDG Metadata Catalogue (MDC).
Because this service is queried by users, contributors (and hopefully
users) are advised to understand QCDml.

After the first working version~\cite{ref:MDWG2004} was released in 2004,
the MDWG has revised the QCDml schema several times~\cite{ref:MDWG2006}. 
Although our design strategy and the global structure of QCDml remain
unchanged, many improvements have been made to meet community
requirements, some of them being incompatible with previous versions. 
In this article, we present key ingredients of the QCDml, based on
the latest release. 

\section{QCDml Structure}
QCDml defines the {\it ensemble XML} and the {\it configuration XML} whose
structure is shown below. The former describes metadata common 
across an ensemble such as <physics> information, while the latter  
contains configuration specific information such as a trajectory
number <update>, a tag <series> which distinguishes different runs
in one ensemble, <precision> of the configuration,
and an <avePlaquette> value. The <algorithm> parameters
can be stored in either of the XML documents.
The <management> part includes information of 
who and when data and metadata are submitted or modified. 
The <implementation> part contains information of machine and code used
in simulations.

The two XML IDs and the configuration itself are linked together in the
following way.
The Markov Chain URI (MCU) in <markovChainURI>
is an unique ensemble name and appears in both XML
documents. The Logical File Name (LFN) in <dataLFN> of the configuration
XML is an unique configuration name and is embedded in the
configuration file~\cite{ref:FileFormat}. The format of the MCU (LFN) is 
``mc:(lfn:)\//\//(Name of Regional Grid)\//(Regional Grid Dependent String)``.

Root elements, <markovChain> for the ensemble XML and <gaugeConfiguration>
for the configuration XML, contain names of namespaces (xmlns=) and
URL's of schema (schemaLocation). A two digit version number of 
the QCDml (currently 1.4 for ensemble and 1.3 for configuration) 
is appended at the end of the namespace, and more one digit release
number is added to the schemata file to identify backward compatible
updates. 

{\small
\ \\
{\it \underline{ensemble XML}}
\begin{verbatim}
<markovChain xmlns:xsi="http://www.w3.org/2001/XMLSchema-instance" 
  xsi:schemaLocation="http://www.lqcd.org/ildg/QCDml/ensemble1.4 
     http://www.lqcd.org/ildg/QCDml/ensemble1.4/QCDmlEnsemble1.4.1.xsd" 
     xmlns="http://www.lqcd.org/ildg/QCDml/ensemble1.4">
  <markovChainURI>
     mc://JLDG/CP-PACS+JLQCD/RCNF2+1/RC28x56_B2050Kud013560
  </markovChainURI>  
  <management/> 
  <physics/> 
  <algorithm/> 
</markovChain> 
\end{verbatim}

\noindent
{\it \underline{configuration XML}}
\begin{verbatim}
<gaugeConfiguration xmlns:xsi="http://www.w3.org/2001/XMLSchema-instance"
   xsi:schemaLocation="http://www.lqcd.org/ildg/QCDml/config1.3
      http://www.lqcd.org/ildg/QCDml/config1.3/QCDmlConfig1.3.0.xsd"
      xmlns="http://www.lqcd.org/ildg/QCDml/config1.3"> 
  <management/> 
  <implementation/>
  <algorithm/>  
  <precision>double</precision> 
  <markovStep>
    <markovChainURI>
      mc://JLDG/CP-PACS+JLQCD/RCNF2+1/RC28x56_B2050Kud013560
    </markovChainURI> 
    <series>2</series>  <update>002200</update> 
    <avePlaquette>6.04326596981212e-01</avePlaquette> 
    <dataLFN>
      lfn://JLDG/CP-PACS+JLQCD/RCNF2+1/RC28x56_B2050Kud013560-2-002200
    </dataLFN> 
  </markovStep>
</gaugeConfiguration>
\end{verbatim}
}

\section{Physics Part}
\subsection{Structure}
The <physics> part of the ensemble XML consists of <size>, <action> 
and optional <observables>. The <size> section looks like
<size> <elem> <name>X</name> <length>16</length> </elem> 
(lengths in Y,Z,T directions) </size>. 

\begin{figure}[t]
\begin{center}
\includegraphics[scale=0.49]{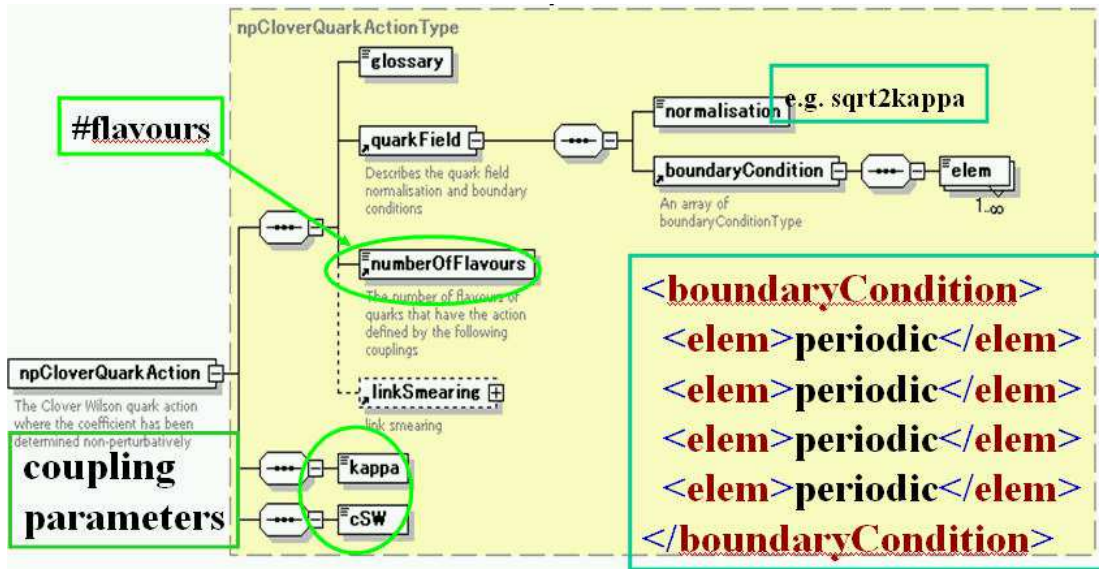}
\caption{A graphical representation of the XML schemata for the clover
quark action.}
\label{fig:action}
\end{center}
\end{figure}

The <action> section is a central part of the ensemble XML. 
The example below illustrates the structure:
{\small
\begin{verbatim}
  <action> 
    <gluon> <iwasakiRGGluonAction/> </gluon> 
    <quark> <npCloverQuarkAction/> <npCloverQuarkAction/> </quark>
  </action>
\end{verbatim}
}
\noindent
Each lattice action is described in an element block
(e.g. <iwasakiRGGluonAction/>) whose name is easily understood.
This strategy is taken, firstly
to assign an unique name to each action, and secondly to realize a
hierarchal structure of action trees in the schemata.
Note that we repeat quark action blocks if coupling parameters
are different, e.g. for the $N_f=2+1$ case.
Fig.~\ref{fig:action} shows a typical quark action.
An action block consists of
\begin{itemize}
\item <glossary>, the URL of a document prepared by contributors
using either text, pdf, ps, tex, or xml formats. 
The document describes full details of the action.
\item gluon/quark field information, <gaugeField> or <quarkField>,
which includes <boundaryCondition> and supplemental information 
such as <representation> and <normalisation>.
\item <numberOfFlavours> for quark actions.
\item optional <linkSmearing> block which is explained below.
\item list of coupling parameters together with tadpole factors
for tadpole improved actions. 
\end{itemize}

The last optional block <observables> of the <physics> part displays
values of measured physical quantities. 
Each element of the list has the following structure: 
<elem> <name/> <value/> <err/> <glossary/> </elem>.
Currently, <name> of <observables> is one of ampi ($am_\pi$),
amrho ($am_\rho$), mpi\_mrho ($m_\pi/m_\rho$), ar0 ($ar_0$) and
ar1 ($ar_1$). 
QCDml supports only dimensionless quantities, because dimensional
quantities (e.g. lattice spacing) cannot be regarded as properties of
an ensemble. 

\subsection{Details of actions}
The current version of QCDml supports the following lattice actions
(lower camel case convention).
\begin{itemize}
\item gluon actions: 
plaquetteGuonAction, iwasakiRGGluonAction, DBW2GluonAction, 
LuscherWeiszGluonAction, treeLevelSymanzikGluonAction,
tpLuscherWeiszGluonAction, anisotropicWilsonGluonAction,
anisotropicTpWilsonGluonAction
\item quark actions:
KSquarkAction, asqTadQuarkAction, 
wilsonQuarkAction, cloverQuarkAction, tpCloverQuarkAction,
npCloverQuarkAction, wilsonTMQuarkAction, domainWallQuarkAction, 
fatLinkIrrelevantCloverQuarkAction,
anisotropicWilsonQuarkAction,
anisotropicCloverQuarkAction
\end{itemize}
\noindent
A list of coupling names for each action can be found 
in Ref.~\cite{ref:ILDGJP}. 

\begin{figure}[t]
\begin{minipage}{72mm}
\begin{center}
\includegraphics[scale=0.60]{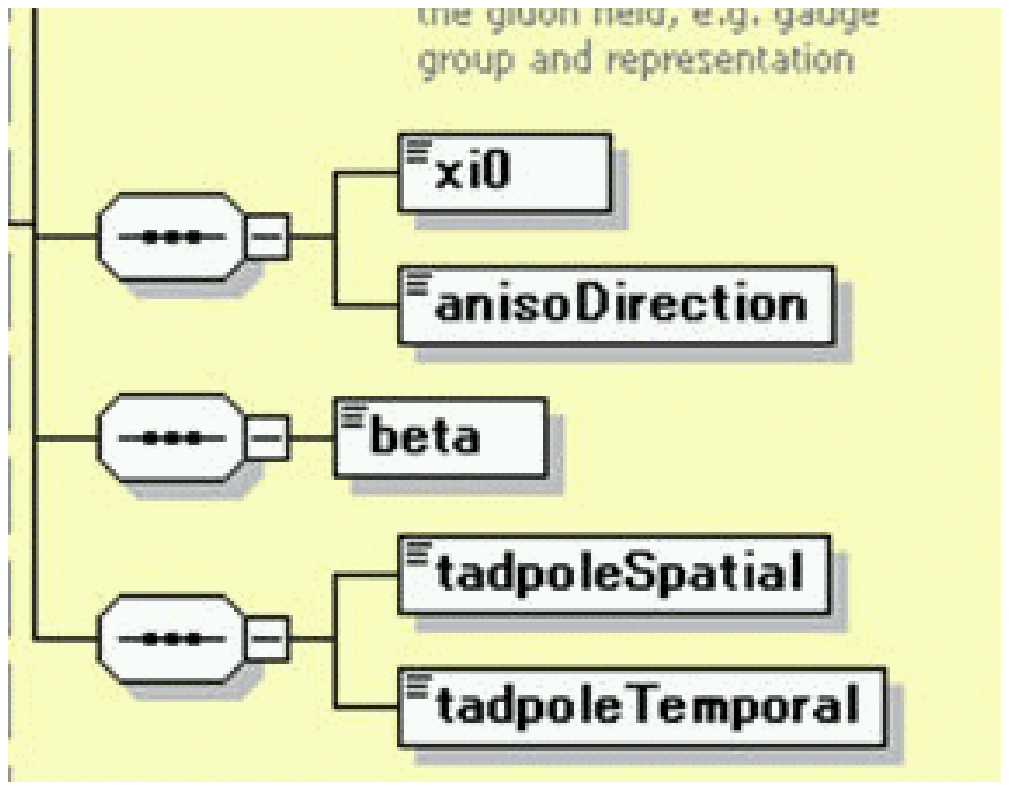}
\caption{Coupling parameters of the anisotropic tadpole improved Wilson
gluon action. <xi0> is the bare gauge anisotropy and <anisoDirection> 
is the direction of anisotropy.}
\label{fig:anisogluon}
\end{center}
\end{minipage}\hspace{4mm}
\begin{minipage}{72mm}
\begin{center}
\includegraphics[scale=0.60]{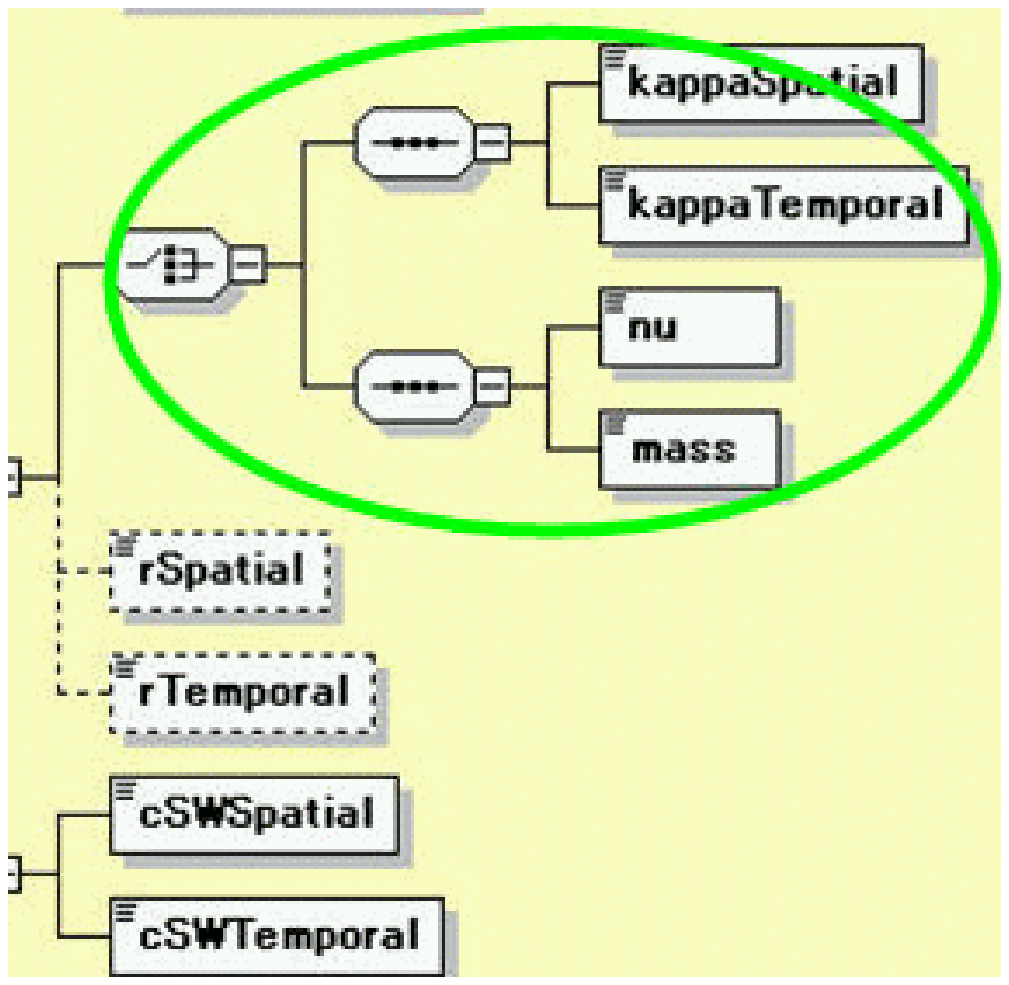}
\caption{Coupling parameters of the anisotropic clover quark action.
Wilson parameters are optional.}
\label{fig:anisoquark}
\end{center}
\end{minipage}
\end{figure}

There are currently two examples of non-unique notation in QCDml.
\begin{itemize}
\item
Gluon actions with plaquette and six-link loops (sixLinkGluonAction),
such as \\<DBW2GluonAction>, are mathematically written as 
\[
S=\beta\times\left(c_0 \sum{\rm plaq.} 
                 + c_1 \sum{\rm rect.} 
                 + c_2 \sum{\rm parallelogram}
                 + c_3 \sum{\rm char}\right). 
\]
The QCDml supports two standard normalisation of couplings;
\[
   c_0=1, \hspace{1cm} c_0 + 8c_1 + 8c_2 + 16c_3 = 1.
\]
The normalisation is marked up as a value of the <normalisation> element,
c0\_is\_one for the former (recommended for tadpole improved actions)
and cs\_sum\_to\_one for the latter (recommended for others).
\item The current version of QCDml supports some anisotropic actions.
See figs.~\ref{fig:anisogluon} and \ref{fig:anisoquark} for examples.
Note that one can choose either 
$(\kappa_{\rm Spatial},\kappa_{\rm Temporal})$ or
$(\mu,{\rm mass})$ notations for anisotropic wilson/clover quark 
actions.
\end{itemize}

\subsection{Link smearing}
Smearing link variables is becoming one of the standard technologies
for improvement. QCDml supports marking up smearing
as an optional <linkSmearing> block in all quark actions.
Link smearing consists of a blocking of link variables using e.g. 
an APE type blocking 
\[
  U_\mu(n) \leftarrow \rho_0 U_\mu(n)
   + \rho_1\sum_{\nu (\ne\mu)} \big[
      U_\nu(n)U_\mu(n+\hat\nu)U^\dagger_\nu(n+\hat\mu)
    + U^\dagger_\nu(n-\hat\nu)U_\mu(n-\hat\nu)U_\nu(n-\hat\nu+\hat\mu) \big]
\]
and a projection onto the SU(3) group (unitarization) using e.g. 
a stout procedure. 
One repeats the set of ``blocking'' and ``unitarization'' several times. 
QCDml therefore marks up link smearing procedures as\\
{\small
\begin{verbatim}
<linkSmearing> 
  <apeLinkBlocking> <rho0/> <rho1/> </apeLinkBlocking> 
  <stoutLinkUnitarization/>
  <numSmear/>
</linkSmearing> 
\end{verbatim}
}
\noindent
In addition to the above example, we currently support 
<anisotropicApeLinkBlocking> and <invSqrtLinkUnitarization>.
The strategy of inserting an optional <linkSmearing> block to
all actions makes it easy to markup various fat link actions. 

\section{Algorithm Part}
Marking up algorithms is realized with a strategy different from
that for actions.
Due to the variety and complexity of algorithms, a generic
hierarchical markup is difficult. Moreover, it is unlikely that
the algorithm would form the primary search criteria.
Therefore, algorithms are marked up in a rather unconstrained way.

The algorithm part in the ensemble XML looks like 
{\small <algorithm> <name/> <glossary/> <reference/> <exact/>
<parameters/> </algorithm>}.
The name of the algorithm is given by contributors as a value of
the <name> element. <glossary> and <reference> elements, also
unconstrained, refer to the URL of a document and an article reference,
respectively. The only constrained element <exact> provides
information of whether the algorithm is exact (true) or not (false).
The <parameters> block contains a list of <name> and <value> pairs like
{
\small
\begin{verbatim}
<parameters>
  <name>PHMC_Polynomial_Order</name>   <value>250</value>
  <name>PHMC_Polynomial_Type</name>    <value>Chebyshev</value>
  <name>HMC_MD_dt</name>               <value>0.400000000E-02</value>
  <name>HMC_MD_steps</name>            <value>250</value>
</parameters>
\end{verbatim}
}

Algorithm parameters might be changed when generating an ensemble.
For example, the molecular dynamics step size of HMC is often tuned in 
a run. Contributors can mark up such configuration dependent
parameters in the <algorithm> part of the configuration XML.
 
In addition to the <parameters> list, contributors can import a
collaboration specific namespace and mark up algorithm parameters
in a hierarchical way. Information given in this way is 
utilized only within the collaboration. 

\section{Management Part}
The <management> part of the ensemble XML describes information for 
the XML document itself. This part includes <collaboration>,
<projectName>, optional <ensembleLabel> and <reference>.
These elements provide users with a starting point to search 
useful ensembles.
The <management> part of the configuration XML includes <crcCheckSum>
which is a crc check sum of the binary data part of the configuration
file~\cite{ref:FileFormat}.
Checking this value ensures that the configuration is successfully
transfered from the ILDG.

In addition, we can record operations
(add, replace, remove) affecting an ensemble XML document in the MDC
together with the time stamp and the participant responsible 
for the operation. 
Similarly, operations (generate, add, replace, remove) affecting
a configuration (or configuration XML metadata) are recorded in 
the <management> part of the configuration XML. 
Marking up these operations is optional, except for 
the information when a configuration has been generated.

\section{Summary and Future Work}
At present, interoperability of the regional grids has been
achieved~\cite{ref:MWWG2007} for download operations
and valuable configurations have already been 
archived~\cite{ref:ILDG2007} in the grid.
Because the ILDG is becoming a new research infrastructure 
for the lattice QCD community, the MDWG encourages our colleagues to
mark up their ensembles and configurations using QCDml. 

The description of QCDml given in this report is not complete.
To study more about QCDml, please visit the ILDG MDWG
web site~\cite{ref:ILDGJP} where you can find sample XML documents
together with human readable documents generated from annotations
embedded in the QCDml schema. 

The lattice actions used in simulations are becoming more and more
complicated. We plan to prepare human readable documents including 
mathematical expressions and reference articles in the near future.
We are going to update QCDml continuously to meet community
requirements.\\[0.2cm] 
%
% \section{Acknowledgments}
We are grateful to all members of the ILDG Metadata Working Group,
in particular to C.~DeTar for helpful suggestions on the manuscript.
T.Y. is partly supported by the Grant-in-Aid of the Ministry of Education 
(No. 18104005 % ukawa, kiban-S
) of the Japanese Government.

\end{document}